\documentclass[%
reprint,%
aps,%
prl,%
groupedaddress,%
superscriptaddress,%
amssymb,%
amsmath,%
floatfix%
]{revtex4-1}

\usepackage{graphicx}
\usepackage[
  colorlinks=true,
  urlcolor=blue,
  linkcolor=blue,
  citecolor=blue
]{hyperref}
\usepackage{siunitx}
\usepackage{braket}

\begin{document}
\title{Microcavity-based Generation of Full Poincar\'{e} Beams with Arbitrary Skyrmion Numbers}
\author{Wenbo Lin}
\email{lin-w@iis.u-tokyo.ac.jp}
\affiliation{
Research Center for Advanced Science and Technology, The University of Tokyo, 4-6-1 Komaba, Meguro-ku, Tokyo 153-8505, Japan
}
\affiliation{
Institute of Industrial Science, The University of Tokyo, 4-6-1 Komaba, Meguro-ku, Tokyo 153-8505, Japan
}
\author{Yasutomo Ota}
\affiliation{
Institute for Nano Quantum Information Electronics, The University of Tokyo, 4-6-1 Komaba, Meguro-ku, Tokyo 153-8505, Japan
}
\author{Yasuhiko Arakawa}
\affiliation{
Institute for Nano Quantum Information Electronics, The University of Tokyo, 4-6-1 Komaba, Meguro-ku, Tokyo 153-8505, Japan
}
\author{Satoshi Iwamoto}
\affiliation{
Research Center for Advanced Science and Technology, The University of Tokyo, 4-6-1 Komaba, Meguro-ku, Tokyo 153-8505, Japan
}
\affiliation{
Institute of Industrial Science, The University of Tokyo, 4-6-1 Komaba, Meguro-ku, Tokyo 153-8505, Japan
}
\affiliation{
Institute for Nano Quantum Information Electronics, The University of Tokyo, 4-6-1 Komaba, Meguro-ku, Tokyo 153-8505, Japan
}

\date{\today}

\begin{abstract}
A full Poincar\'e (FP) beam possesses all possible optical spin states in its cross-section, which constitutes an optical analogue of a skyrmion. Until now, FP beams have been exclusively generated using bulk optics. Here, we propose a generation scheme of an FP beam based on an optical microring cavity. We position two different angular gratings along with chiral lines on a microring cavity and generate an FP beam as a superposition of two light beams with controlled spin and orbital angular momenta. We numerically show that FP beams with tailored skyrmion numbers can be generated from this device, opening a route for developing compact light sources with unique optical spin fields.
\end{abstract}


\maketitle
Spin angular momentum (SAM; $s$) and orbital angular momentum (OAM; $l$) are fundamental parameters characterizing photons~\cite{allen1992orbital,simpson1997mechanical} in terms of the polarization state and a helical wavefront, respectively. Recent progress in the control of SAM and OAM in optical beams has led to the generation of various vector beams
with textured optical spin fields~\cite{milione2011higher,yi2015hybrid,naidoo2016controlled,liu2017generation}. Among them, full Poincar\'e (FP) beams exhibit extremely distinctive photonic spin distributions: an FP beam possesses all possible photonic spin states in its beam cross-section~\cite{beckley2010full}. This unique optical spin field is receiving significant attention with potential for various applications, such as novel polarization sensing~\cite{suarez2019mueller} and optical tweezers~\cite{wang2012optical,zhu2015transverse}. It is known that FP beams can be generated by superposing two light beams with opposite SAM states and different OAM states. Until now, this method has been examined using bulk optics, resulting in the generation of Laguerre--Gaussian~\cite{beckley2010full,ling2016characterization,wang2017generation,galvez2012poincare} and Bessel-type~\cite{shvedov2015visualizing,lopez2019overall} FP beams.

A notable property of FP beams is that their optical spin textures constitute skyrmions in two dimensions~\cite{donati2016twist,gao2019skyrmionic}.
A skyrmion is a topological elementary excitation characterized by a topological number or a skyrmion number, $N_\mathrm{sk}$.
Skyrmions have been observed in various condensed-matter systems, such as chiral magnetic materials~\cite{yu2010real}, Bose--Einstein condensates~\cite{choi2012observation}, and chiral liquid crystals~\cite{fukuda2011quasi}. Optical skyrmions and their derivatives have been observed in some optical fields, including confined modes in nanophotonic structures~\cite{cilibrizzi2016half,tsesses2018optical,du2019deep,bai2020dynamic,guo2019meron,wang2019observation} as well as in FP beams~\cite{donati2016twist,watzel2020topological}. In conjunction with the concept of optical skyrmion, FP beams offer an interesting area  for exploring rich physics, such as the transfer of topological charges between light and matter~\cite{donati2016twist} and the formation of photonic M\"{o}bius strips~\cite{bauer2015observation}. 

Until now, FP beams have been exclusively generated using bulk optics, which is not suitable for practical applications that require robustness and compactness of light sources. For unlocking the potential of FP beams in various applications and physics, it is highly desired to develop FP beam generators on chip.

Here, we propose a scheme for generating an FP beam based on an optical microring cavity. Our approach utilizes a tightly confined whispering-galley mode (WGM), in which the spin--orbit interaction of light provides polarization singularities~\cite{luxmoore2013interfacing,sollner2015deterministic,coles2016chirality}, enabling SAM-controlled light diffraction transferring the OAM into free space by the tailored perturbation of the structure. We perturb the cavity by patterning two types of angular diffraction gratings~\cite{cai2012integrated,xiao2016generation,shao2018spin}. Accordingly, an FP beam is synthesized as a superposition of two diffracted beams with controlled SAM and OAM in the far field. We analytically show that this method in principle can produce FP beams with arbitrary $N_\mathrm{sk}$, and numerically demonstrate the generation of FP beams with various $N_\mathrm{sk}$s. Our scheme could facilitate the integration of FP beam sources with different $N_\mathrm{sk}$s on a single chip, which may widen the application of FP beams.

\begin{figure}[t]
    \centering
    \includegraphics{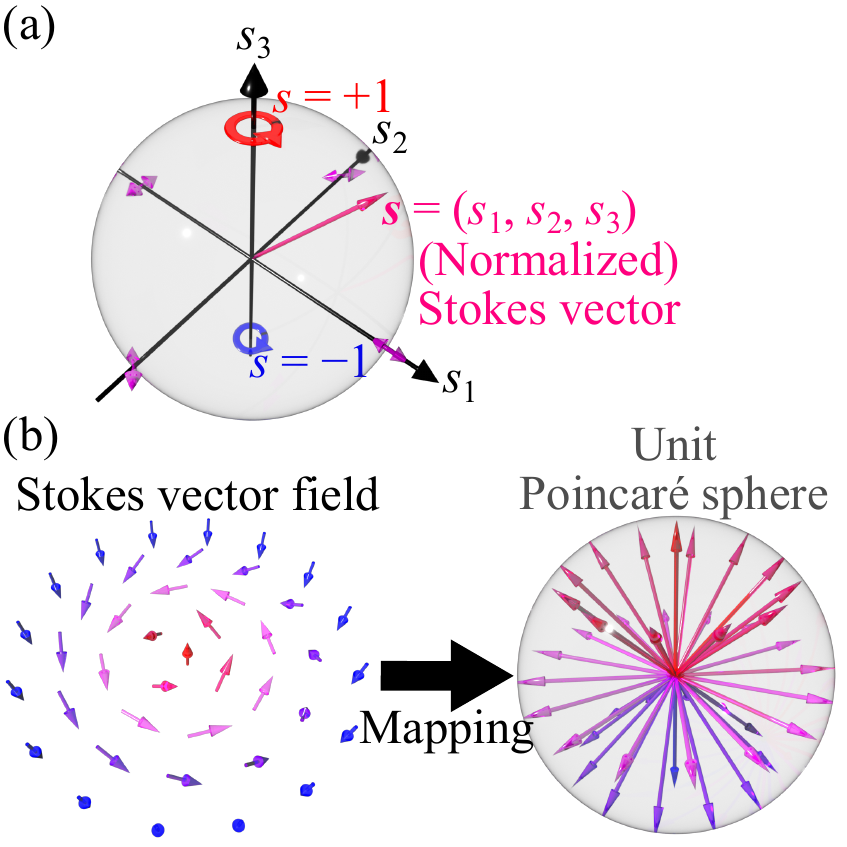}
    \caption{%
    (a) Unit Poincar\'e sphere defined in a space spanned by the normalized Stokes parameters, which represent the degree of linear polarization ($s_1$), diagonal polarization ($s_2$), and circular polarization ($s_3$), respectively. A vector from the origin to a point on the sphere surface defines a Stokes vector and indicates the orientation of the optical spin.
    (b) Schematic of a cross-sectional Stokes vector field of an example FP beam and its projection to a unit Poincar\'e sphere.
    }
    \label{fig:PoincareSphere&StokesVevtor}
\end{figure}

An arbitrary polarization state or a spin state of a photon can be described using a normalized Stokes vector composed of three parameters: $\boldsymbol{s} = \left(S_1,S_2,S_3\right)/S_0$. Each parameter corresponds to an expectation value of a Pauli matrix   for a photonic spin state~\cite{martinelli2017polarization}.
$\boldsymbol{s}$ denotes the orientation of a photonic spin, which can be visualized by a vector arrow drawn in a unit Poincar\'e sphere, as shown in Fig.~\ref{fig:PoincareSphere&StokesVevtor}(a). Figure~\ref{fig:PoincareSphere&StokesVevtor}(b) schematically depicts a cross-sectional Stokes vector field of an example FP beam and its projection to the surface of a unit Poincar\'e sphere. An FP beam possesses any possible spin states in the cross-section, and hence, projecting them to the surface completely wraps the sphere.
The topological property or the order of an FP beam is characterized by a skyrmion number that counts the number of times the photonic spins in a certain area ($A$) wrap a unit Poincar\'e sphere, which is expressed as
\begin{equation}
    N_\mathrm{sk} = \frac{1}{4\pi}\int_A{\boldsymbol{s} \cdot \left[ \partial_x \boldsymbol{s} \times \partial_y \boldsymbol{s} \right]\mathrm{d}x\mathrm{d}y}.
    \label{eq:skyrmionNumber}
\end{equation}
For an FP beam generated by superposing two beams with opposite SAMs ($\boldsymbol{s}_2 = -\boldsymbol{s}_1$) and different absolute OAMs ($ \left| l_2 \right| \neq \left| l_1 \right| $), Eq.~(\ref{eq:skyrmionNumber}) yields the OAM difference, i.e., $N_\mathrm{sk} = \left( l_2 - l_1 \right) = \Delta{l}$. This expression is obtained by integrating Eq.~(\ref{eq:skyrmionNumber}) for a domain $A$ where the spin state of the superposed beam flips from $\boldsymbol{s}_1$ to $-\boldsymbol{s}_1$ (derivation can be found in Supplemental Material (SM)).
Therefore, FP beams with any skyrmion numbers can be generated by controlling the OAMs of the beams under superposition~\cite{ling2016characterization,wang2017generation}. The capability of generating skyrmions with $N_\mathrm{sk}$s much larger than 1 could be of interest, because  the extensively investigated magnetic skyrmions frequently exhibit $N_\mathrm{sk}$ values of only 1 or 2~\cite{ozawa2017zero,zhang2016high}.

\begin{figure*}[t]
    \centering
    \includegraphics{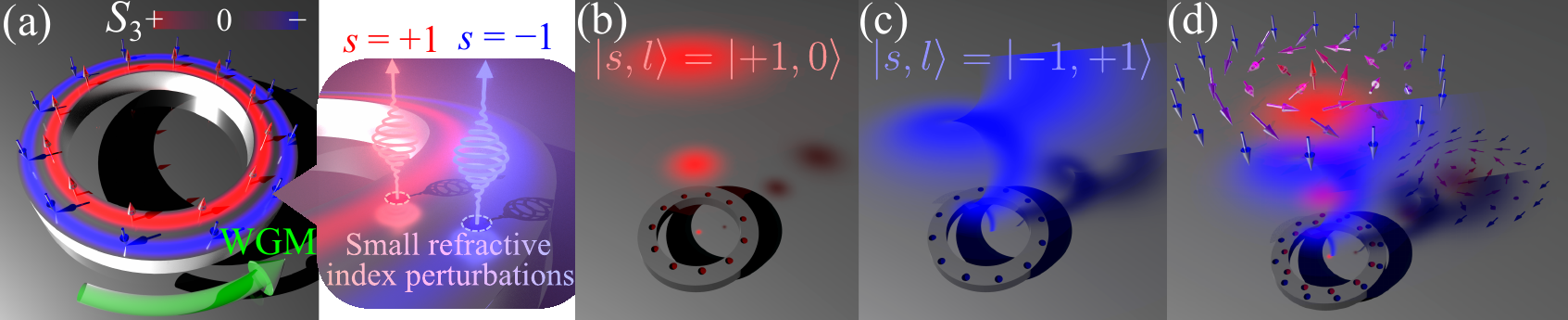}
    \caption{%
    (a)~Schematic showing a photonic spin distribution for a counterclockwise WGM in a micoring cavity. The region with high positive (negative) spin density is indicated by red (blue). There are two singular lines with pure spin polarization of $s = +1$ or $-1$, called as $C$-lines. The inset shows a zoomed image of the microring, presenting light scattering due to the refractive index perturbations on the $C$-lines. The scattered light is circularly polarized, and its polarity corresponds to that of the perturbed $C$-line.
    (b)~Light diffraction from a ring with an angular grating of $g = m-1$ on the $C$-line of $s = +1$. The diffracted light is in an angular momentum state of $\ket{s,l} = \ket{+1,0}$.
    (c)~Same as (b) but with a grating of $g = m$ positioned on the $C$-line of $s = -1$. The diffracted light is in an angular momentum state of $\ket{s,l} = \ket{-1,+1}$.
    (d)~Synthesis of the two diffracted light beams in (b) and (c) by simultaneously arranging the two gratings on the microring, resulting in the generation of an FP beam with an $N_\mathrm{sk}$ of 1.
    }
    \label{fig:Concept}
\end{figure*}
Now, we discuss our scheme of the microcavity-based generation of FP beams. We consider a microring cavity supporting a WGM rotating within the cavity. The tight spatial confinement for the propagating mode induces the spin--orbit interaction of light~\cite{luxmoore2013interfacing,sollner2015deterministic,coles2016chirality}, and thus leads to the coupling between th e spin and orbit degrees of freedom within the mode. Consequently, the WGM can be described by a superposition of spin-up and spin-down components $\psi_\pm$ with angular momentum states of $\ket{s,l} = \ket{\pm{1},m\mp{1}}$, where $m$ is the azimuthal order of the WGM and yields the total angular momentum on multiplication with $\hbar$.
Figure~\ref{fig:Concept}(a) schematically presents  the distribution of the net spin density ($S_3 = \left| \psi_+ \right|^2 - \left| \psi_- \right|^2$) of a counterclockwise transverse electric (TE)-like WGM in a microring cavity.
The spatial profiles of the spin-up and spin-down modes differ, resulting in the emergence of purely spin polarized lines, known as $C$-lines~\cite{yu2015exceptional,bauer2016optical,garcia2017optical}.
We utilize this phenomenon for achieving SAM-controlled light scattering from the WGM. We apply a small refractive index perturbation aligned to a $C$-line, which selectively scatters circularly polarized photons, with the handedness depending on the polarity of the $C$-line, as schematically shown in the right inset in Fig.~\ref{fig:Concept}(a). 

By arranging a circular, periodic array of index perturbations, we can control the OAM of the scattered light according to the angular momentum conservation law expressed as  $(m-s)-l = ng$~\cite{cai2012integrated,xiao2016generation}. Here, $m-s$ is the effective OAM of the WGM that the angular grating feels, $l$ is the OAM of the diffracted light, $n$ is the diffraction order, and $g$ is the number of grating elements. Hereafter, we focus on the first-order diffraction of $n = 1$. Figures~\ref{fig:Concept}(b) and (c) show example behaviors of the microcavity with gratings. When an angular grating with $g = m-1$ is patterned on the $C$-line of $s = +1$, a light beam described by $\ket{s,l} = \ket{+1,0}$ will be generated (Fig.~\ref{fig:Concept}(b)). Meanwhile, positioning a grating of $g = m$ on the $C$-line of $s = -1$ generates a beam with $\ket{s,l} = \ket{-1,+1}$ (Fig.~\ref{fig:Concept}(c)).

By arranging the above two gratings in parallel, an FP beam can be synthesized in far field as a superposition of the light beams diffracted by the two grati ngs.
The far-field amplitude profile, $A{\left( \theta, \phi \right)}$, of an FP beam generated by our scheme can be analytically approximated by
\begin{equation}
A{\left( \theta, \phi \right)} = J_{\left|l\right|}{\left( k d_0 \sin{\theta} \right)}\exp\left[\mathrm{i}l\left( \phi - \frac{\pi}{2} \right)\right],
\label{eq:farFieldProfile}
\end{equation}
where $J_{\left|l\right|}$ is the $|l|$th-order Bessel function of the first kind, $k$ is the wavenumber, $d_0$ is the radius of the circle where the angular grating is patterned, $\phi$ is the azimuth angle, and $\theta$ is the elevation angle ($\theta = 0$ corresponds to the direction normal to the device plane).
This analytical expression was derived with a toy model that assumes each grating element acts as a small electric dipole (see SM and Refs.~\cite{ostrovsky2013generation,vaity2015perfect}).

\begin{figure}[tb]
    \centering
    \includegraphics{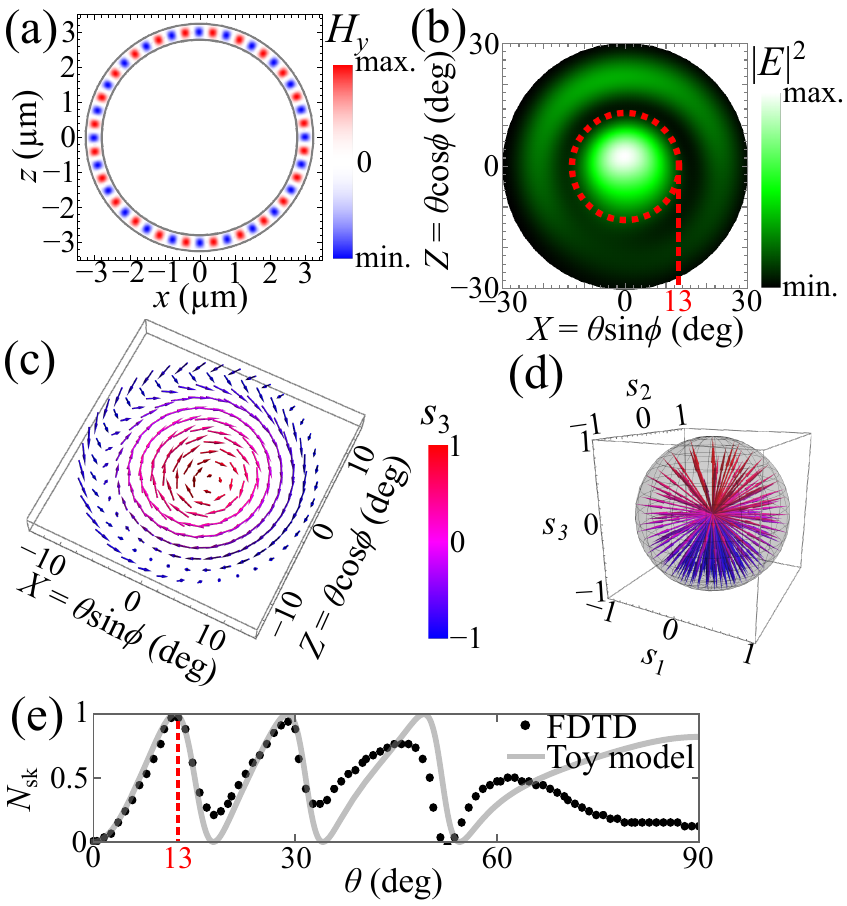}
    \caption{%
    (a)~Calculated mode profile of a counterclockwise TE-like WGM with an azimuthal order of 24. The out-of-plane magnetic field is shown in the figure.
    (b)~Projected far-field intensity when two angular gratings of $g = 23$ and $24$ are applied along the $C$-lines of $s = +1$ and $s = -1$, respectively.
    Here, we considered the collimation of the far field and its projection to a flat plane. The conversion to the Cartesian coordinates in the flat plane ($X,Z$) can be achieved by multiplying an appropriate scale factor to the coordinates defined using the elevation ($\theta$) and the azimuth ($\phi$) angles.
    (c)~Stokes vector field within $\theta = 13$\si{\degree}.
    (d)~Stokes vector of (c) projected to a unit Poincar\'e sphere.
    (e)~Skyrmion number $N_\mathrm{sk}$ versus elevation angle $\theta$.
    }
    \label{fig:basicNsk}
\end{figure}

As a specific design of an FP beam generator, we consider a silicon microring cavity with a radius of 3~\si{\micro\meter}. 
The waveguide of the cavity has a width of 450~\si{\nano\meter} and a height of 200~\si{\nano\meter}, confining a single TE mode at the telecommunication C-band. 
The refractive indices of silicon and the background environment were set to 3.4 and 1.0, respectively. 
The ring cavity supports a WGM of an azimuthal order of 24 around a wavelength of 1.57~\si{\micro\meter} with a large free spectral range of $\approx 30$~\si{\nano\meter}.
This mode has a high $Q$ factor of $\approx 3 \times 10^8$ without any index perturbation.
We focus on the counterclockwise WGM and analyze it using finite-difference time-domain (FDTD) simulations. A calculation domain of $3.5\times3.5\times1.0$~\si{\micro\meter^3} and a grid size of $10\times10\times10$~\si{\nano\meter^3} were employed in the simulation. We selectively excited the mode in the simulator and continued the computation until the stationary state was reached.
Figure~\ref{fig:basicNsk}(a) shows a computed mode profile of the unperturbed micoring cavity.
We found $C$-lines of $s = +1$ and $s = -1$ at the positions respectively deviated by $-0.14$~\si{\micro\meter} and $+0.125$~\si{\micro\meter} from the waveguide center in the radial direction. Along each $C$-line, we arranged an array of air holes with a radius and a depth of both 40~\si{\nano\meter}. The far fields radiated from the structure were calculated from the obtained near fields using the near-to-far field conversion technique.

First, we designed an FP beam generator emitting a beam with $N_\mathrm{sk} = +1$. We patterned 23 holes on the $C$-line of $s = +1$ and 24 holes on the $C$-line of $s = -1$, so that the beams in the states of $\ket{s,l} = \ket{+1,0}$ and $\ket{s,l} = \ket{-1,+1}$ will be respectively diffracted and superposed.
Figure~\ref{fig:basicNsk}(b) presents the intensity profile of the calculated far field within an elevation angle of $\theta \leq 30$\si{\degree} projected onto a flat plane after collimating the beam virtually.
The red dashed line in Fig.~\ref{fig:basicNsk}(b) indicates $\theta$ of 13\si{\degree}.
The values of $\theta = $13\si{\degree} and 30\si{\degree} approximately correspond to the first and second minimums in the intensity profile of the light diffracted only by the angular grating of $s = +1$ (see SM).
Thus, the directions of the spins are expected to be aligned downward ($s \approx -1$) at these angles.
Figure~\ref{fig:basicNsk}(c) shows the  spatial distribution of the optical spin field within $\theta \leq 13$\si{\degree}. Spins are directed upward near the center, rotate as changing azimuth angle, and gradually flip their direction to downward as the domain edge is approached. Figure~\ref{fig:basicNsk}(d) plots a projection of the spin texture to a unit Poincar\'e sphere. The full coverage of the surface demonstrates that the generated beam is indeed an FP beam. The projection map suggests that the observed optical spin field forms a Bloch-type skyrmion. We note that other types of skyrmions, such as N\'eel-type one, can be generated by rotating one of the angular gratings with respect to the other (not shown).

Subsequently, we calculated $N_\mathrm{sk}$ according to Eq.~(\ref{eq:skyrmionNumber}).
The integration was performed from the center to a particular $\theta$ and for all $\phi$.
Figure~\ref{fig:basicNsk}(e) displays the $\theta$ dependence of $N_\mathrm{sk}$.
At $\theta \approx 13$\si{\degree}, $N_\mathrm{sk}$ reaches a near unity value of 0.98, which compares well with the designed value of $N_\mathrm{sk} = +1$. 
The remaining error of 0.02 may be caused by the imperfect circular polarization of the scattered light, resulting from the finite size of the scatter and/or from the multiple reflection of the scattered light before exiting the structure.
We also analytically deduced values of $N_\mathrm{sk}$ using our toy model and overlaid the result on the plot in Fig.~\ref{fig:basicNsk}(e). The analytical curve explains well the FDTD results, in particular when $\theta$ is small, in which case the toy model can be regarded as a good approximation.
For both the FDTD and analytical results, we observed oscillations in $N_\mathrm{sk}$ for $\theta > 13$\si{\degree}. The maxima and minima of the oscillations correspond to the point where the intensity of one of the two scattered beams under superposition approaches zero. Analytically, $N_\mathrm{sk}$ should oscillate between 0 and the designed $N_\mathrm{sk}$, leading to the generation of a skyrmion and an antiskyrmion alternately.
This skyrmionic structure is known as a skyrmion multiplex~\cite{fujita2017ultrafast}, the simplest case of which is known as skyrmionium. Another skyrmion state with $N_\mathrm{sk} = 1/2$, called as half skyrmion or meron~\cite{yu2018transformation}, could also be produced by spatially filtering the beam by a diaphragm passing only $\theta < 8$\si{\degree}.

\begin{figure}[tb]
    \centering
    \includegraphics{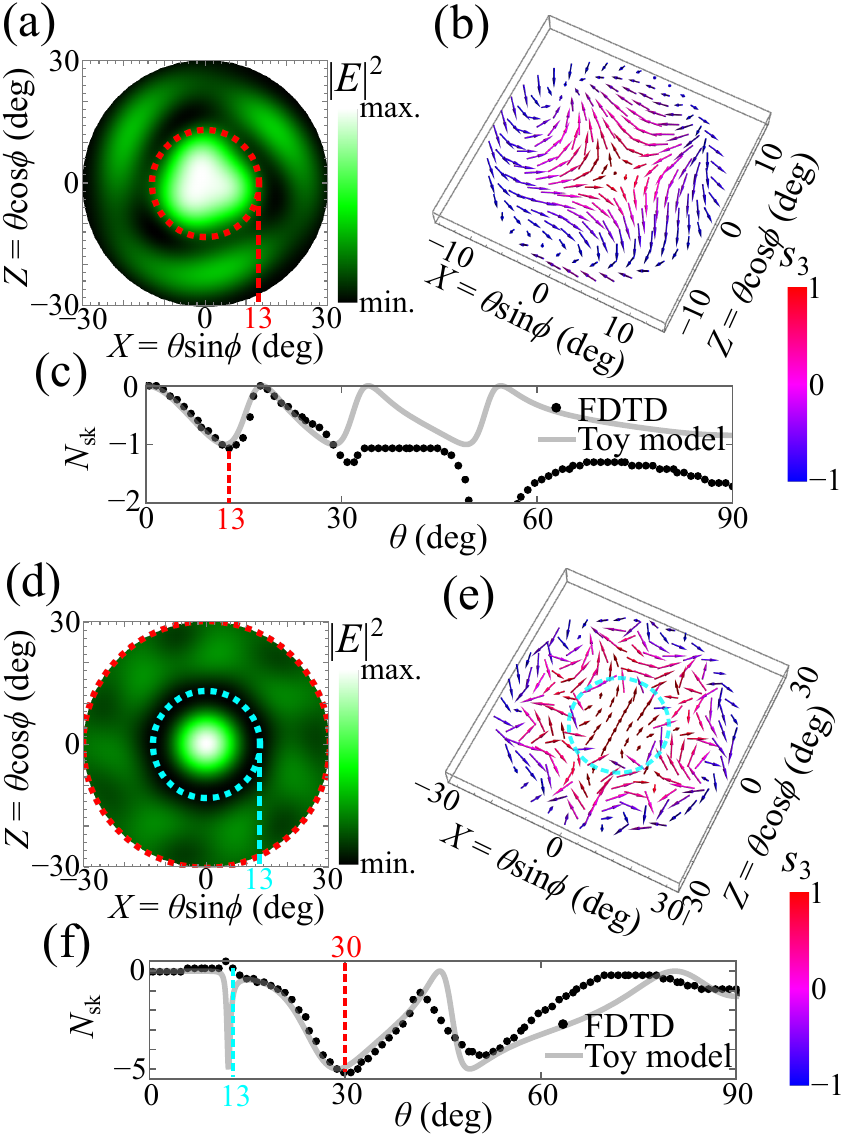}
    \caption{%
    Generation of FP beams with different $N_\mathrm{sk}$s.
    (a--c) Beam properties of the device designed for generating an FP beam with $N_\mathrm{sk}$ of $-1$; (a) Projected far-field intensity profile, (b) Stokes vector field, and (c) calculated skyrmion number versus elevation angle.
    (d--f) Same as (a--c) but for the device designed with $N_\mathrm{sk}$ of $-5$.
    }
    \label{fig:variousNsk}
\end{figure}
Next, we synthesized an FP beam with $N_\mathrm{sk} = -1$ or an antiskyrmionic beam. 
For this, $g$ of the grating on the $C$-line of $s = -1$ was changed from 24 to 26, so that a beam in a  state of $\ket{s,l} = \ket{-1,-1}$ will be radiated from the grating.
Figures~\ref{fig:variousNsk}(a) and (b) respectively show the  computed far-field intensity profile and the corresponding optical spin field within $\theta = 13$\si{\degree}. The latter texture shows an antivortex behavior, as expected. The calculated $N_\mathrm{sk}$ reaches $-1.07$ at $\theta = 13$\si{\degree}, as presented in Fig.~\ref{fig:variousNsk}(c). These results demonstrate that an antiskyrmionic FP beam can also be generated within our scheme.
Finally, we present an example device generating an FP beam with a large $N_\mathrm{sk}$ of $-5$.
For this demonstration, $g$ of the grating on the $C$-line of $s = -1$ was increased to 30, which radiatesa beam in the state of $\ket{s,l} = \ket{-1,-5}$.
Figures~\ref{fig:variousNsk}(d) and (e) show the  computed far-field intensity profile and the corresponding spin field, respectively.
From the latter plot, it can be observed that the optical spins wind five times around the field center. 
Figure~\ref{fig:variousNsk}(f) displays the $\theta$ dependence of the calculated $N_\mathrm{sk}$. The minimum $N_\mathrm{sk}$ of $-5.22$ was found at $\theta = 30$\si{\degree}, demonstrating an FP beam with a large $N_\mathrm{sk}$ of $-5$.
Meanwhile, for the toy model, $N_\mathrm{sk}$ should reach $-5$ close to $\theta = 13$\si{\degree} (indicated by cyan broken lines in Figs.~\ref{fig:variousNsk}(d)--(f)).
However, we did not observe an $N_\mathrm{sk}$ of approximately $-5$ at $\theta = 13$\si{\degree} in the FDTD simulations. This unexpected behavior in the computation model seemingly arises from the weak diffracted beam in the state of $\ket{s,l} = \ket{-1,-5}$ close to $\theta = 13$\si{\degree}. The weak signal is obsecured by the background noise  in the simulator, and thus, yields an unpredictable result. Here, the large $l$ of the beam induces the weak signal at small $\theta$ values. More robust generation of FP beams with high $N_\mathrm{sk}$s could be possible by engineering the size and/or shape of the microring and the pattern of the far fields.

In summary, we demonstrated FP beam generation from a microring cavity. We augmented the microring by angular gratings patterned on the $C$-lines, which diffracted SAM-controlled beams with different OAMs. We showed that the superposition of the beams from the microcavity results in an FP beam with an arbitrary controllable $N_\mathrm{sk}$. We examined concrete designs of silicon-based microring cavities and verified the generation of FP beams with various $N_\mathrm{sk}$s.
We note that the clarified Bessel-like beam patterns of the generated FP beams may lead to their self-healing propagation~\cite{vyas2011self,li2017generation,otte2018recovery}. 
We believe that the compact FP beam generators could find broad applications in optical communication and optical sensing and play a significant role in exploring condensed matter physics.

\begin{acknowledgements}
This research was supported by JST CREST Grant Number JPMJCR19T1, JSPS KAKENHI Grant Number JP15H05700, JP17H02796, and JP19J13955.
\end{acknowledgements}

\bibliography{ms}

\end{document}


\title{Supplemental Materials for\\``\textit{Microcavity-based Generation of Full Poincar\'{e} Beams with Arbitrary Skyrmion Numbers}''}
\author{Wenbo Lin}
\email{lin-w@iis.u-tokyo.ac.jp}
\affiliation{
Research Center for Advanced Science and Technology, The University of Tokyo, 4-6-1 Komaba, Meguro-ku, Tokyo 153-8505, Japan
}
\affiliation{
Institute of Industrial Science, The University of Tokyo, 4-6-1 Komaba, Meguro-ku, Tokyo 153-8505, Japan
}
\author{Yasutomo Ota}
\affiliation{
Institute for Nano Quantum Information Electronics, The University of Tokyo, 4-6-1 Komaba, Meguro-ku, Tokyo 153-8505, Japan
}
\author{Yasuhiko Arakawa}
\affiliation{
Institute for Nano Quantum Information Electronics, The University of Tokyo, 4-6-1 Komaba, Meguro-ku, Tokyo 153-8505, Japan
}
\author{Satoshi Iwamoto}
\affiliation{
Research Center for Advanced Science and Technology, The University of Tokyo, 4-6-1 Komaba, Meguro-ku, Tokyo 153-8505, Japan
}
\affiliation{
Institute of Industrial Science, The University of Tokyo, 4-6-1 Komaba, Meguro-ku, Tokyo 153-8505, Japan
}
\affiliation{
Institute for Nano Quantum Information Electronics, The University of Tokyo, 4-6-1 Komaba, Meguro-ku, Tokyo 153-8505, Japan
}
\date{\today}
\maketitle

\section{A toy model for describing diffraction patterns}
We develop a toy model consisting of a circular array of radiative dipoles for analytically deriving diffraction patterns of angular diffraction gratings interacting with a whispering-galley mode (WGM). We consider infinitesimal dipoles, the respective polarization of which emulates the local optical polarization of the WGM. We assumed that other scattering mechanisms occurring in the microring cavity are negligible and solely considered the radiation from the dipoles located equidistantly. The radiation from one of the dipoles can be written using a well-known approximated form given by
\begin{equation}
    \boldsymbol{E}_\mathrm{far,single}{\left( \boldsymbol{r} \right)} \propto \frac{\boldsymbol{r} - \boldsymbol{r_0}}{\left| \boldsymbol{r} - \boldsymbol{r_0} \right|} \times \left[ \frac{\boldsymbol{r} - \boldsymbol{r_0}}{\left| \boldsymbol{r} - \boldsymbol{r_0} \right|} \times \boldsymbol{E}_\mathrm{WGM}{\left( \boldsymbol{r}_0 \right)} \right] \frac{\mathrm{e}^{\mathrm{i}k\left| \boldsymbol{r} - \boldsymbol{r_0} \right|}}{\left| \boldsymbol{r} - \boldsymbol{r_0} \right|},
    \label{eq:DipoleRadiation}
\end{equation}
where $\boldsymbol{r}_0$ is the position of the dipole and $k$ is the wavenumber. The source dipole moment is replaced by a local electric field vector of the WGM,  $\boldsymbol{E}_\mathrm{WGM}{\left( \boldsymbol{r}_0 \right)}$, so as to express the scattered radiation of the mode by a grating element at $\boldsymbol{r}_0$.
In far-field region ($kr \gg 1$ and $r \gg r_0$), Eq.~(\ref{eq:DipoleRadiation}) can be further approximated to a spherical wave:
\begin{eqnarray}
    \boldsymbol{E}_\mathrm{far,single}{\left( \boldsymbol{r} \right)}
    & \sim & \frac{\boldsymbol{r} \times \left[ \boldsymbol{r} \times \boldsymbol{E}_\mathrm{WGM}{\left( \boldsymbol{r}_0 \right)} \right]}{r^2} \frac{\mathrm{e}^{\mathrm{i}kr}}{r} \exp{\left[ -\mathrm{i}k \frac{\boldsymbol{r}\cdot\boldsymbol{r}_0}{r} \right]} \nonumber \\
    & = & \boldsymbol{P}_\mathrm{far,single}{(\phi,\theta)}R{(r)}\tilde{E}_\mathrm{far,single}{(\phi,\theta)}.
    \label{eq:DipoleRadiation_far}
\end{eqnarray}
In the last equal of this equation, we decomposed the far field into three components. The first factor, $\boldsymbol{P}_\mathrm{far,single}{(\phi,\theta)} = \boldsymbol{r} \times \left[ \boldsymbol{r} \times \boldsymbol{E}_\mathrm{WGM} \right]/r^2$, represents the vectorial nature of the field.
The set of the cross products, $\boldsymbol{r}\times\boldsymbol{r}\times$, projects the polarization of the source dipole field (which equals $\boldsymbol{E}_\mathrm{WGM}$) to the plane normal to the wavevector (or to the vector $\boldsymbol{r}$).
The second factor, $R{(r)} = \mathrm{e}^{\mathrm{i}kr}/r$, expresses the propagation of a scalar spherical wave. The third factor,  $ \tilde{E}_\mathrm{far,single}{(\phi,\theta)} = \exp{\left[ -\mathrm{i}k \boldsymbol{r}\cdot\boldsymbol{r}_0/r \right]} $, is also a scalar part of the far field. The last two factors merely describe the field intensity distribution in the far field and do not scramble the light polarization. Therefore, the polarization distribution can be solely represented by $\boldsymbol{P}_\mathrm{far,single}{(\phi,\theta)}$.
We evaluate the vector part or the light polarization on a coordinate described by
\begin{equation}
    \left( \begin{array}{c}
        E_R\\
        E_X\\
        E_Y
    \end{array} \right)
    =
    \left( \begin{array}{ccc}
        1 & 0 & 0\\
        0 & \cos\phi & -\sin\phi\\
        0 & \sin\phi & \cos\phi
    \end{array} \right)
    \left( \begin{array}{c}
        E_r\\
        E_\theta\\
        E_\phi
    \end{array} \right).
    \label{eq:ProjectedFarFieldCoordinate}
\end{equation}
$E_{X}$ and $E_{Y}$ components correspond to the light polarization in the collimated far field. Then, $\boldsymbol{P}_\mathrm{far,single}$ can be expressed as
\begin{equation}
    \boldsymbol{P}_\mathrm{far,single}{(\phi,\theta)}
    =
    -\left( \begin{array}{c}
        0\\
        E_{\mathrm{WGM},x}\\
        E_{\mathrm{WGM},y}
    \end{array} \right)
    +
    \left( \begin{array}{c}
        E_{\mathrm{WGM},x}\\
        E_{\mathrm{WGM},y}\\
        E_{\mathrm{WGM},z}
    \end{array} \right)
    \cdot
    \left( \begin{array}{c}
        \left( 1-\cos\theta \right)\cos{\phi}\\
        \left( 1-\cos\theta \right)\sin{\phi}\\
        \sin\theta
    \end{array} \right)
    \left( \begin{array}{c}
        0\\
        \cos{\phi}\\
        \sin{\phi}
    \end{array} \right).
    \label{eq:PolarizationProjection}
\end{equation}
The second term in the right hand side becomes negligible as far as both $E_z$ and $\theta$ are small. These conditions match with our case, where light scattering from a TE mode ($E_z \approx 0$) is considered mainly for $\theta$ of no more than 30\si{\degree}. 
We note that, even at $\theta = 30$\si{\degree}, radiation from a circular dipole exhibit a high degree of circular polarization of $|s_3| = |S_3 / S_0| = -2\operatorname{Im}{\left[ {E_Y}^\ast E_X \right]} / \left( {E_X}^\ast E_X + {E_Y}^\ast E_Y \right) \approx 0.99$ in the far field, well sustaining the one-to-one correspondence between the polarization of a local source dipole and that of the collimated far field. These discussions render an approximated equation, $\boldsymbol{P}_\mathrm{far,single} \sim -(0,E_\mathrm{WGM,x},E_\mathrm{WGM,y})$ for small $\theta$.

Now, we discuss far field created by an array of the scattering elements. Considering the phase difference of radiation among the grating elements, the superposed radiation field can be expressed as
\begin{equation}
    \boldsymbol{E}_\mathrm{far}{\left( \boldsymbol{r} \right)} \sim \sum_{j=0}^{g-1}{\frac{\boldsymbol{r} - \boldsymbol{r_i}}{\left| \boldsymbol{r} - \boldsymbol{r_j} \right|} \times \left( \frac{\boldsymbol{r} - \boldsymbol{r_j}}{\left| \boldsymbol{r} - \boldsymbol{r_j} \right|} \times \boldsymbol{E}_{\mathrm{WGM},j} \exp{\left[\mathrm{i}m2\pi\frac{j}{g}\right]} \right) \frac{\mathrm{e}^{\mathrm{i}k\left| \boldsymbol{r} - \boldsymbol{r_j} \right|}}{\left| \boldsymbol{r} - \boldsymbol{r_j} \right|}}.
    \label{eq:DipoleRadiation_AngularGrating}
\end{equation}
Here $g$ is the number of grating elements. $\boldsymbol{r}_j = d_0\left(\cos{\left[ 2\pi j/g \right]},\sin{\left[ 2\pi j/g \right]}, 0 \right)$ is the position of the $j$th grating element. The grating elements are arranged with equal intervals on a circle in the $x$-$y$ plane having a center $(0,0,0)$ and a radius $d_0$.
$\boldsymbol{E}_{\mathrm{WGM},j}$ is the polarization vector of the WGM at $\boldsymbol{r}_j$.
$m$ is the azimuthal order of the WGM.
Following the discussion about the single dipole,
the vector part of Eq.~(\ref{eq:DipoleRadiation_AngularGrating}) also sustains the one-to-one correspondence with the polarization of the local source dipole for small $\theta$. 
This means that the polarization distribution becomes homogeneous within the approximation if all the grating elements scatter light with the same polarization.
Here, we assume that $\boldsymbol{E}_{\mathrm{WGM},j}$ is a circular polarization with a spin angular momentum (SAM) of $s$.
Then, $\boldsymbol{E}_{\mathrm{WGM},j}$ can be rewritten to $\boldsymbol{E}_{\mathrm{WGM},0}\exp{\left[ -\mathrm{i}s2\pi j/g \right]}$, where $\boldsymbol{E}_{\mathrm{WGM,}0}$ is the polarization vector of the WGM at $(d_0, 0, 0)$.
Taking into account this additional phase factor of $\exp{\left[ -\mathrm{i}s2\pi j/g \right]}$, together with $\exp{\left[ \mathrm{i}m2\pi j/g \right]}$ in Eq.~(\ref{eq:DipoleRadiation_AngularGrating}), the $r$-independent scalar part of Eq.~(\ref{eq:DipoleRadiation_AngularGrating}), $\tilde{E}_\mathrm{far}$, can be written as
\begin{equation}
    \tilde{E}_\mathrm{far}{\left( \phi, \theta \right)} \propto 
   \frac{1}{g}\sum_{j=0}^{g-1}{\exp{\left[-\mathrm{i} k \sin{\theta} d_0 \cos{\left( 2\pi\frac{j}{g} - \phi \right)} \right]} \exp{\left[\mathrm{i}(m-s)2\pi\frac{j}{g}\right]}}.
    \label{eq:PolygonalSourcesRadiation}
\end{equation}
This equation takes a form of inverse discrete Fourier transform. We consider the limit of infinite $g$ and obtain a continuous Fourier transform given by
\begin{equation}
    \tilde{E}_\mathrm{far}{\left( \phi, \theta \right)} \xrightarrow{g \rightarrow \infty} \sum_{n=-\infty}^{\infty}{ \frac{1}{2\pi}\int_{0}^{2\pi}{\exp{\left[-\mathrm{i} \rho \cos{\left( \xi - \phi \right)} \right]} \exp{\left[ \mathrm{i} l_n \xi \right]} \mathrm{d}\xi}},
    \label{eq:PolygonalSourcesRadiation_Limits}
\end{equation}
where $\rho = k \sin{\theta} d_0$. $l_n = m - s - ng$ is the orbital angular momentum (OAM) order of the diffracted light of $n$th order diffraction. The integral in Eq.~(\ref{eq:PolygonalSourcesRadiation_Limits}) constitutes an inverse Fourier transform of a so-called perfect optical vortex with an OAM of $l_n$, and is known to be a $l_n$th order Bessel function of the first kind, $J_{l_n}(\rho)$~\cite{ostrovsky2013generation,vaity2015perfect}.
Indeed, we can show that Eq.~(\ref{eq:PolygonalSourcesRadiation_Limits}) can be interpreted as a Fourier transform of the generating function of $J_{l_n}(\rho)$. The generating function is expressed as
\begin{equation}
    \exp{\left[ \frac{\rho}{2}\left( t - t^{-1} \right) \right]} = \sum_{l_n = -\infty}^{\infty}J_{l_n}{\left( \rho \right)}t^{l_n}.
    \label{eq:BesselJ_GeneratingFunciton}
\end{equation}
By replacing $t \rightarrow \mathrm{e}^{\mathrm{i}\xi}$, we get a Fourier series
\begin{equation}
    \exp{\left[ -\mathrm{i}\rho\sin{\xi} \right]} = \sum_{l_n = -\infty}^{\infty}J_{\left|l_n\right|}{\left( \rho \right)}\mathrm{e}^{\mathrm{i}\xi l_n},
    \label{eq:FourierSerise_BesselJ}
\end{equation}
which yields
\begin{equation}
    \frac{1}{2\pi}\int_{0}^{2\pi}{\exp{\left[-\mathrm{i} \rho \sin{\xi} \right]} \exp{\left[\mathrm{i}l_n\xi\right]} \mathrm{d}\xi} = J_{\left|l_n\right|}{\left( \rho \right)}.
    \label{eq:FourierTransform_BesselJ}
\end{equation}
This equation has the same form with the integral in Eq.~(\ref{eq:PolygonalSourcesRadiation_Limits}). By Substituting Eq.~(\ref{eq:FourierTransform_BesselJ}) into Eq.~(\ref{eq:PolygonalSourcesRadiation_Limits}), we finally obtain
\begin{equation}
    \tilde{E}_\mathrm{far}{\left( \phi, \theta \right)} \propto \sum_{n=-\infty}^{\infty}{J_{|l_n|}{\left( k d_0 \sin{\theta} \right)}\exp{\left[ \mathrm{i}l_n\left( \phi - \frac{\pi}{2} \right)\right]}}.
    \label{eq:FarField_BesselApprox}
\end{equation}
This equation gives a good approximation to Eq.~(\ref{eq:PolygonalSourcesRadiation}) when the ``\textit{spatial sampling rate}'' or $g$ is large and when the ``\textit{spatial frequency}'' or $\theta$ is small. These conditions also meet with the situation discussed in the main text, where the first order diffraction is treated and the dominant radiation is concentrated to the region with small $\theta$ as far as $m \approx g$. 

\section{Detailed FDTD results}
\begin{figure*}[b]
    \centering
    \includegraphics{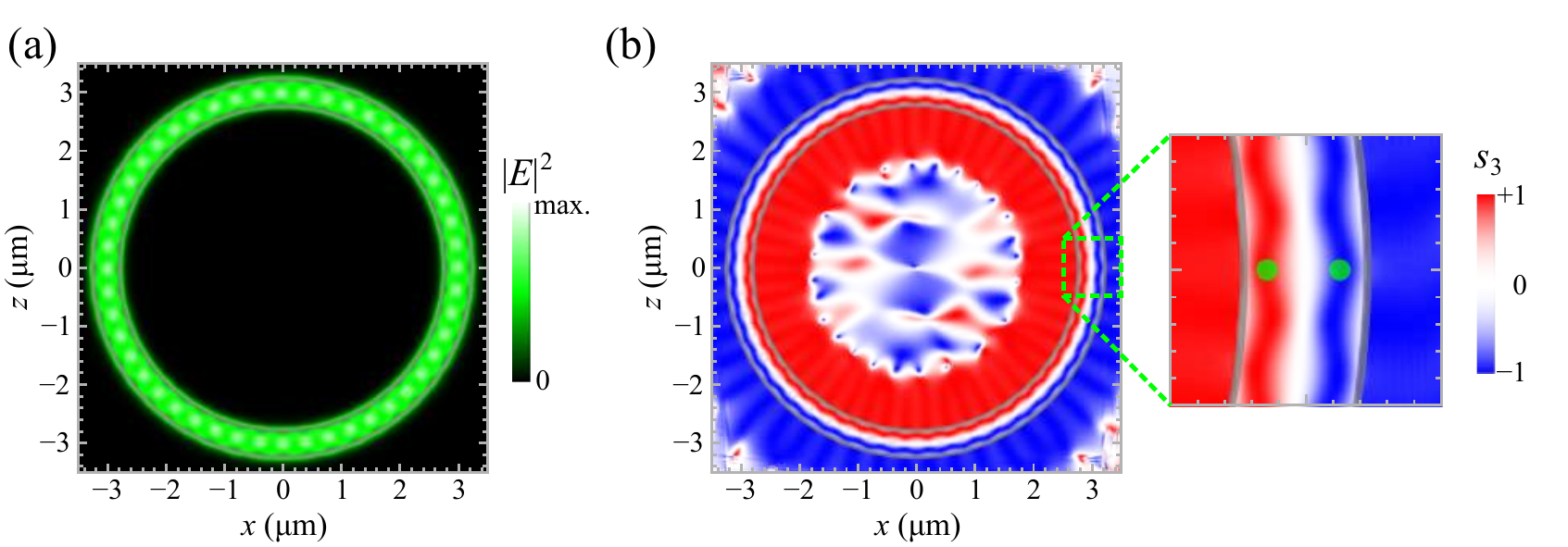}
    \caption{%
    Calculated counterclockwise WGM of the azimuthal order of 24 for the unperturbed microring cavity:
    (a)~Field intensity profile, (b)~Spatial distribution of the degree of circular polarization, which equals to the normalized Stokes parameter $s_3$. The right inset is a magnified view of the area surrounded by the green dashed line. The green dots in the inset show the radial positions and dimensions of the grating elements on the $C$-lines.
    }
    \label{fig:CavityMode}
\end{figure*}
In this section, we discuss more details of the results computed by the finite-difference time-domain (FDTD) method.
Figures~\ref{fig:CavityMode}(a) and (b) respectively show the calculated intensity distribution and photonic spin density distribution ($s_3 = -2\operatorname{Im}{\left[ {E_z}^\ast E_x \right]} / \left( {E_x}^\ast E_x + {E_z}^\ast E_z \right)$) of a WGM without gratings. Here, we computed a TE-like counterclockwise-rotating WGM with the azimuthal order of 24.
The intensity distribution is nearly constant in the azimuthal direction but with slight modulation. The wavy pattern is induced by the intrusion of the counter-rotating WGM, which was weakly excited in the simulator due to the imperfect mode selection. The similar distribution can be found in the spin density distribution, which shows two wavy lines with mutually opposite polarity. 
The spin-up (-down) region is found near the inner (outer) edge of the microring, which is much clearly displayed in the inset of Fig.~\ref{fig:CavityMode}(b). The light-green dots show the radial positions of the grating elements. They are placed above the $C$-lines.

\begin{figure*}[tb]
    \centering
    \includegraphics{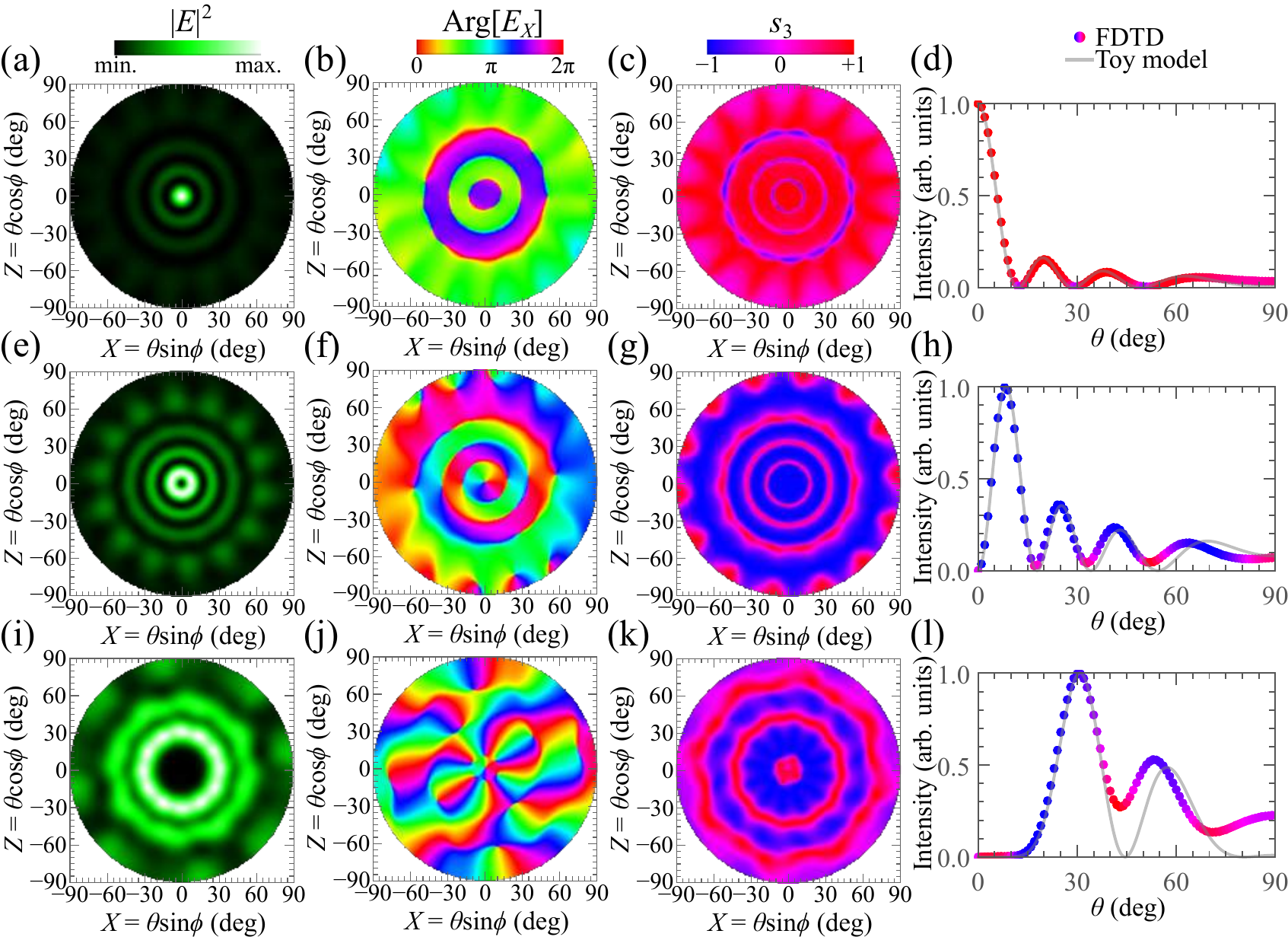}
    \caption{%
    (a--d)~Far field of the light diffracted by the angular grating with 23 grating elements on the $s=+1$ $C$-line: (a)~Projected intensity profile, (b)~Phase distribution of the projected electric field $E_X$, (c)~Photonic spin density distribution, (d)~Field intensity plot along with the line of $Z = 0$. The color of the dot reflects the local spin state. The gray solid line shows the corresponding result obtained by the toy model.
    (e--h)~Same as (a--d) but computed for the device with an angular grating of $g = 24$ on the $C$-line of $s=-1$.
    (i--l)~Same as (e--h) but with a grating of $g = 30$.
    }
    \label{fig:FDTDvsToyModel}
    \hrule height.5pt
\end{figure*}
We computed far fields based on the near-to-far field conversion method, using near fields recorded at a plane just above the microring cavity.
Figures~\ref{fig:FDTDvsToyModel}(a--d) show computed far-field patters for a perturbed ring cavity with a 23-element grating on the $C$-line of $s=+1$, which is designed to diffract a beam in a state of $\ket{s,l}=\ket{+1,0}$.
The intensity profile plotted in Fig.~\ref{fig:FDTDvsToyModel}(a) exhibits a circularly-symmetric beam pattern. 
Figure~\ref{fig:FDTDvsToyModel}(b) shows a phase profile of the diffracted beam.
In the azimuth direction, the phase is almost constant, which corresponds to an OAM state of $l = 0$.
Figure~\ref{fig:FDTDvsToyModel}(c) shows the spatial distribution of the spin density distribution ($s_3 = -2\operatorname{Im}{\left[ {E_Z}^\ast E_X \right]} / \left( {E_X}^\ast E_X + {E_Z}^\ast E_Z \right)$). Nearly pure and homogeneous spin-up polarization is observed across the entire far field, suggesting the successful radiation of circularly polarized light from the $C$-line. The minor blue regions, where the spin state is not pure spin-up anymore, occur in the places where the far-field intensity itself is very weak. We consider that, at these regions in the far field, unspecified background noise in the simulator causes the unexpected spin flip.
Figure~\ref{fig:FDTDvsToyModel}(d) shows a slice of the far field along with the line of $Z = 0$. For comparison, we also plot the result obtained by the toy model. The two lines results agree well within the range of $\theta \leq 30$\si{\degree}.

Figures~\ref{fig:FDTDvsToyModel}(e--h) show the same set of results with those depicted in (a--d) but of the device with a 24-element grating on the $C$-line of $s=-1$, which is designed to diffract a beam in a state of $\ket{s,l}=\ket{-1,+1}$.
The intensity plot shows a doughnut-like distribution around the far field center.
Together with this result, the phase distribution plot demonstrates that the far field carry OAM of $l=-1$. The spin density distribution shows that the dominant spin state is $s=-1$. Again, the unwanted spin flip occurs in the region with weak far-field intensity, which is more visible in the current case. Nevertheless, we confirmed a good agreement between the FDTD and the toy models as in Fig.~\ref{fig:FDTDvsToyModel}(h).

The same calculations were also performed for the device with a 30-element grating on the $C$-line of $s=-1$, as plotted in Figs.~\ref{fig:FDTDvsToyModel}(e--h). In this case, far field in the state of $\ket{s,l}=\ket{-1,-5}$ will be diffracted from the cavity.
The intensity plot exhibits an even wider doughnuts pattern in the far field, which was confirmed to carry OAM of $l=-5$ in conjunction with the plot of the phase pattern. The spin distribution exhibit the unwanted spin flip even near the far-field center ($\theta \approx 0$\si{\degree}). This unexpected result arises from further weaker field intensity around the field center owing to the widen hole in the doughnut beam with the increased $|l|$. Even in this situation, the FDTD result agrees well with the toy mode when $\theta < 30$\si{\degree}.

\section{Analytical skyrmion number for an FP beam}
Here, we analytically derive an expression for a skyrmion number of an FP beam generated by superposing two vortex beams with different OAM and opposite SAM values. We use a Jones vector $\ket{s_1}$ for expressing a polarization state. $\ket{s_1}$ is defined in a $x$-$y$ plane and can be fully described with two parameters: an amplitude ratio ($\tan{\alpha}$) and an phase difference ($\delta$) between the $x$ and $y$ electric field components. Then, $\ket{s_1}$ is expressed as
\begin{equation}
    \ket{s_1} = \left( \begin{array}{c} \cos{\alpha} \\ \mathrm{e}^{\mathrm{i}\delta}\sin{\alpha}  \end{array} \right).
    \label{eq:JonesVector1}
\end{equation}
The opposite (or orthogonal) polarization state $\ket{s_2}$ is written as
\begin{equation}
    \ket{s_2} = \left( \begin{array}{c} -\sin{\alpha} \\ \mathrm{e}^{\mathrm{i}\delta}\cos{\alpha}  \end{array} \right),
    \label{eq:JonesVector2}
\end{equation}
which satisfies $\Braket{s_2 | s_1} = 0$.
The polarization state $\ket{s_1}$ can also be described using a three-component Stokes vector
\begin{equation}
    \boldsymbol{s}_1 = \left( \begin{array}{c} \cos{2\alpha} \\ \sin{2\alpha}\cos{\delta} \\ \sin{2\alpha}\sin{\delta} \end{array} \right),
    \label{eq:StokesVector1}
\end{equation}
and the Stokes vector for the orthogonal state is $\boldsymbol{s}_2 = -\boldsymbol{s}_1$.
Meanwhile, the other angular momentum, orbital angular momentum (OAM), can be characterized by a spiral wavefront. It is known that the mode cross section of an optical beam carrying a well-defined OAM is circularly symmetric in general, and the spiral wavefront can be written as
\begin{equation}
    \psi{(r,\phi)} = A{(r)} \mathrm{e}^{\mathrm{i}\left[ l \phi + \gamma{(r)} \right]},
    \label{eq:WavefrontOfOAM}
\end{equation}
where $r$ is the radial distance from the beam center, $\phi$ is the azimuth angle, $A$ is the amplitude, $l$ is the order of OAM, and $\gamma$ is the phase offset.

Accordingly, the polarization state of an FP beam generated as the superposition of two different OAM beams with opposite SAMs can be written as
\begin{equation}
    \boldsymbol{\psi}{(r,\phi)} = A_1{(r)} \mathrm{e}^{\mathrm{i}\left[ l_1 \phi + \gamma_1{(r)} \right]} \left( \begin{array}{c} \cos{\alpha} \\ \mathrm{e}^{\mathrm{i}\delta}\sin{\alpha}  \end{array} \right) + A_2{(r)} \mathrm{e}^{\mathrm{i}\left[ l_2 \phi + \gamma_2{(r)} \right]} \left( \begin{array}{c} -\sin{\alpha} \\ \mathrm{e}^{\mathrm{i}\delta}\cos{\alpha}  \end{array} \right),
    \label{eq:SuperpositionWavefunction}
\end{equation}
and the normalized Stokes vector deduced from $\boldsymbol{\psi}$ is given by
\begin{equation}
    \boldsymbol{s} = \frac{{A_1}^2-{A_2}^2}{{A_1}^2+{A_2}^2} \boldsymbol{s}_1 + \frac{2 A_1 A_2}{{A_1}^2+{A_2}^2} \left( \begin{array}{c}
        \sin{2\alpha}\cos{\left(\Delta{l}\phi + \Delta{\gamma}\right)} \\
        \cos{2\alpha}\cos{\left(\Delta{l}\phi + \Delta{\gamma}\right)}\cos{\delta} - \sin{\left(\Delta{l}\phi + \Delta{\gamma}\right)}\sin{\delta} \\
        \cos{2\alpha}\cos{\left(\Delta{l}\phi + \Delta{\gamma}\right)}\sin{\delta} + \sin{\left(\Delta{l}\phi + \Delta{\gamma}\right)}\cos{\delta}
    \end{array} \right),
    \label{eq:SuperpositionStokesVector}
\end{equation}
where $\Delta{l} = l_2 - l_1$ and $\Delta{\gamma}{(r)} = \gamma_2{(r)} - \gamma_1{(r)}$.
The skyrmion number of a certain region $A = \left\{(r,\phi) \mid r_1 \leq r \leq r_2, 0 \leq \phi \leq 2\pi \right\}$ is then given by
\begin{eqnarray}
    \frac{1}{4\pi}\iint_A{\boldsymbol{s} \cdot \left[ \partial_x \boldsymbol{s} \times \partial_y  \boldsymbol{s} \right]\mathrm{d}x\mathrm{d}y} &=& \frac{1}{4\pi}\iint_A{\boldsymbol{s}\cdot\left(\partial_r \boldsymbol{s} \times \partial_\phi \boldsymbol{s} \right)\mathrm{d}r\mathrm{d}\phi}\nonumber\\
    &=& \frac{1}{4\pi}\iint_A{4\Delta{l}\frac{ A_1 A_2 \left[ A_1 \partial_r A_2 - \partial_r A_1 A_2 \right] }{\left[ {A_1}^2+{A_2}^2 \right]^2}\mathrm{d}r\mathrm{d}\phi}\nonumber\\
    &=& \frac{\Delta{l}}{2\pi}\int_0^{2\pi}{\mathrm{d}\phi}\int_{r_1}^{r_2}{\partial_r\frac{-A_1^2}{{A_1}^2+{A_2}^2}\mathrm{d}r}\nonumber\\
    &=& \Delta{l}\left[ \frac{A_1{(r_1)}^2}{A_1{(r_1)}^2+A_2{(r_1)}^2} - \frac{A_1{(r_2)}^2}{A_1{(r_2)}^2+A_2{(r_2)}^2} \right].
    \label{eq:SuperpositioLocalCurvature}
\end{eqnarray}
Therefore, the skyrmion number $N_\mathrm{sk}$ is proportional to the difference in OAM $\Delta{l}$, and exactly equals $\Delta{l}$ when the spin states are $\ket{s_1}$ at $r_1$ (or $A_2(r_1) = 0$) and $\ket{s_2}$ at $r_2$ (or $A_1(r_2) = 0$ ).
$N_\mathrm{sk}$ does not depend on the phase offset $\gamma{(r)}$.
Since the mode profile is known to depend on $|l|$, $r_1$ and $r_2$ satisfying such a condition often exist under the condition of $|l_2| \neq |l_1|$. For analytically calculating the skyrmion number of an FP beam generated by our microring cavity, we assumed the Bessel-like far-field distribution discussed in Eq.~(\ref{eq:FarField_BesselApprox}) and obtained
\begin{equation}
    N_\mathrm{sk}{(\theta)} = \Delta{l}\left[ \frac{ J_{|l_1|}{\left( 0 \right)}^2 }{ J_{|l_1|}{\left( 0 \right)}^2 + J_{|l_2|}{\left( 0 \right)}^2 } - \frac{ J_{|l_1|}{\left( k d_1 \sin{\theta} \right)}^2 }{ J_{|l_1|}{\left( k d_1 \sin{\theta} \right)}^2 + J_{|l_2|}{\left( k d_2 \sin{\theta} \right)}^2 } \right],
    \label{eq:BesselNsk}
\end{equation}
where $d_1(d_2)$ is the radius of the angular grating that diffracts a beam in an angular momentum state of $\ket{s,l}=\ket{+1,l_1}(\ket{-1,l_2})$.

\section{Detailed far-field profile of an FP beam}
\begin{figure*}[tb]
    \centering
    \includegraphics{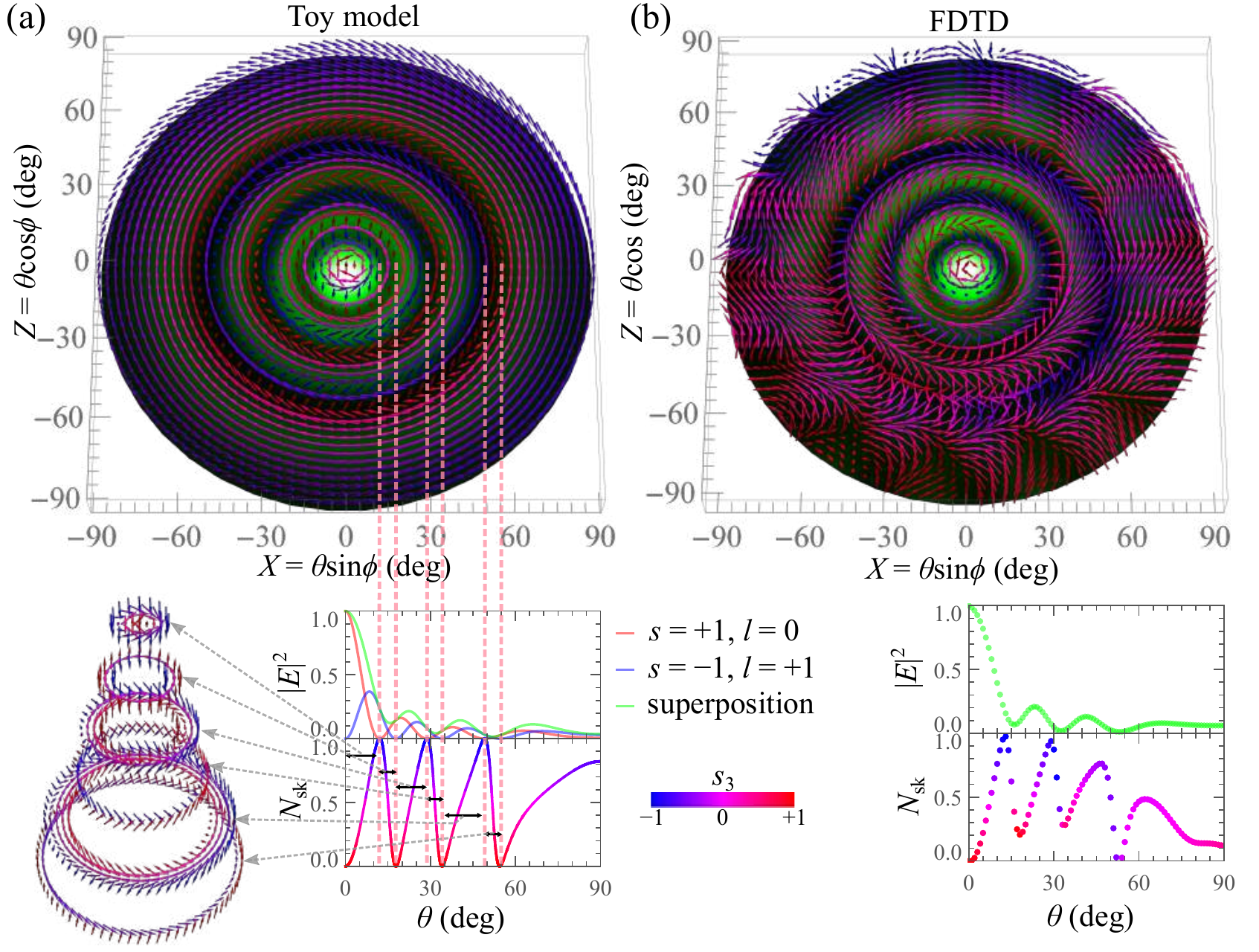}
    \caption{%
    Far-field intensity and spin distribution of an FP beam with $N_\mathrm{sk} = +1$, which is synthesized from two beams in the angular momentum states of $\ket{s,l} = \ket{+1,0}$ and $\ket{-1,+1}$.
    (a) Results by the toy model. The right bottom insets show slices of field intensity and evolution of Nsk. The left bottom inset shows spin distributions between neighbor oscillations of $N_\mathrm{sk}$. Each of the spin distribution constitutes a skyrmion or an antiskyrmion with $|N_\mathrm{sk}| = 1$.
    (b) Results by the FDTD calculations.
    }
    \label{fig:Nsk1_full}
    \hrule height.5pt
\end{figure*}
In this section, we take a closer look to the field pattern of an FP beam. We consider an FP beam with $N_\mathrm{sk} = +1$, which is synthesized from two beams in the angular momentum states of $\ket{s,l} = \ket{+1,0}$ and $\ket{-1,+1}$.
Each beam in the different angular momentum state can be independently generated by solely placing a single angular grating on the microring. The far-field pattern of such a beam generated by the single grating can be described by a Bessel function, as derived in a previous section. 
Figure~\ref{fig:Nsk1_full}(a) shows results obtained by the toy model. The far-field pattern is overlaid with the plot of spin distribution. 
The top panel of the right bottom inset of Fig.~\ref{fig:Nsk1_full}(a) shows far field patterns diffracted from individual angular gratings.
It is obvious that the two beams (colored in red and blue) generated from the different angular gratings have different positions of intensity zero, at which complete spin flips occur in the superposed beam (light green).
The spin flip induce the oscillation of $N_\mathrm{sk}$ as computed in the bottom panel in the right bottom inset of of Figure~\ref{fig:Nsk1_full}(a). The behavior of spin flip is visualized in the left bottom inset. The structure that skyrmions and antiskyrmions appear alternately is knows as skyrmion multiplex~\cite{fujita2017ultrafast}.
In the skyrmionic structure, one can find a variety of skyrmionic structures, such as skyrmionium, skyrmion and meron. These states may be obtained by spatially filtering the FP beam carrying the skyrmion multiplex.

Figure~\ref{fig:Nsk1_full}(b) shows the same set of results calculated using the FDTD method. Overall, the FDTD simulations reproduce the results obtained by the toy model in particular when $\theta$ is small. 
One noticeable deviation can find in the behavior of $N_\mathrm{sk}$, which does not return to zero when varying $\theta$. This is due to the imperfection in the spin flips at the zero field intensity points. In the case of FDTD simulations, unspecified background radiation scrambles the spin state where the diffracted signal is weak.
The behavior of $N_\mathrm{sk}$ also significantly deviates when $\theta$ is large, where the approximation used in the toy model does not stand.
The use of a microring with a larger radius would be helpful for focusing the beam power to the region with small $\theta$ and thereby for generating more ideal FP beams.

\bibliography{sm}